\documentclass[useAMS,usenatbib,referee]{mn2e}
\usepackage[english]{babel}
\usepackage{graphicx}
\usepackage{amssymb, amsmath}

\def\la{\; \raise0.3ex\hbox{$<$\kern-0.75em\raise-1.1ex\hbox{$\sim$}}\;}
\def\ga{\;  \raise0.3ex\hbox{$>$\kern-0.75em\raise-1.1ex\hbox{$\sim$}}\;}
 
\begin{document}

\def\pFn{p_{\raise-0.3ex\hbox{{\scriptsize F$\!$\raise-0.03ex\hbox{\rm n}}}}
}  
\def\pFp{p_{\raise-0.3ex\hbox{{\scriptsize F$\!$\raise-0.03ex\hbox{\rm p}}}}
}  
\def\pFe{p_{\raise-0.3ex\hbox{{\scriptsize F$\!$\raise-0.03ex\hbox{\rm e}}}}
}  
\def\pFmu{p_{\raise-0.3ex\hbox{{\scriptsize F$\!$\raise-0.03ex\hbox{\rm
$\mu$}}}} }  
\def\m@th{\mathsurround=0pt }
\def\eqalign#1{\null\,\vcenter{\openup1\jot \m@th
   \ialign{\strut$\displaystyle{##}$&$\displaystyle{{}##}$\hfil
   \crcr#1\crcr}}\,}
\newcommand{\vp}{\mbox{\boldmath $p$}}         
\newcommand{\vS}{\mbox{\boldmath $S$}}
\newcommand{\vP}{\mbox{\boldmath $P$}}

\newcommand{\vk}{\mbox{\boldmath $k$}}         
\newcommand{\xixi}{\mbox{\boldmath $\xi$}}         
\newcommand{\vq}{\mbox{\boldmath $q$}}         
\newcommand{\vr}{\mbox{\boldmath $r$}}         

\newcommand{\om}{\mbox{$\omega$}}              
\newcommand{\Om}{\mbox{$\Omega$}}              
\newcommand{\Th}{\mbox{$\Theta$}}              
\newcommand{\ph}{\mbox{$\varphi$}}             
\newcommand{\del}{\mbox{$\delta$}}             
\newcommand{\Del}{\mbox{$\Delta$}}             
\newcommand{\lam}{\mbox{$\lambda$}}            
\newcommand{\Lam}{\mbox{$\Lambda$}}            
\newcommand{\ep}{\mbox{$\varepsilon$}}         
\newcommand{\ka}{\mbox{$\kappa$}}              
\newcommand{\dd}{\mbox{d}}                     
\newcommand{\vect}[1]{\bf #1}                
\newcommand{\vtr}[1]{\mbox{\boldmath $#1$}}  
\newcommand{\vF}{\mbox{$v_{\mbox{\raisebox{-0.3ex}{\scriptsize F}}}$}}  
\newcommand{\pF}{\mbox{$p_{\mbox{\raisebox{-0.3ex}{\scriptsize F}}}$}}  
\newcommand{\kF}{\mbox{$k_{\rm F}$}}           
\newcommand{\kTF}{\mbox{$k_{\rm TF}$}}         
\newcommand{\kB}{\mbox{$k_{\rm B}$}}           
\newcommand{\tn}{\mbox{$T_{{\rm c}n}$}}        
\newcommand{\tp}{\mbox{$T_{{\rm c}p}$}}        
\newcommand{\te}{\mbox{$T_{eff}$}}             
\newcommand{\ex}{\mbox{\rm e}}                 
\newcommand{\rate}{\mbox{${\rm erg~cm^{-3}~s^{-1}}$}}
\newcommand{\mur}{\raisebox{0.2ex}{\mbox{\scriptsize (э)}}} 
\newcommand{\Mn}{\raisebox{0.2ex}{\mbox{\scriptsize (э{\it n\/})}}}        %
\newcommand{\Mp}{\raisebox{0.2ex}{\mbox{\scriptsize (э{\it p\/})}}}        %
\newcommand{\MN}{\raisebox{0.2ex}{\mbox{\scriptsize (э{\it N\/})}}}        %

\normalsize
\thispagestyle{empty}



\title[Temperature-dependent pulsations of superfluid neutron stars]
{Temperature-dependent pulsations of superfluid neutron stars}
\author[M. E. Gusakov, N. Andersson]
{M. E. Gusakov$^{1, 2}$\thanks{E-mail:
gusakov@astro.ioffe.ru (MEG); na@maths.soton.ac.uk (NA)}, 
N. Andersson$^{2}$
\\
$^{1}$Ioffe Physical Technical Institute, 
Politekhnicheskaya 26, St.-Petersburg 194021, Russia \\
$^{2}$ School of Mathematics, University of Southampton,
Southampton S017 1BJ, United Kingdom}

\date{Accepted 2006 August 22. Received 2006 August 21; 
in original form 2006 February 13}

\pagerange{\pageref{firstpage}--\pageref{lastpage}} \pubyear{2006}

\maketitle

\label{firstpage}

\begin{abstract}
We examine radial oscillations of
superfluid neutron stars at finite internal temperatures.
For this purpose we generalize
the description of relativistic
superfluid hydrodynamics
to the case of superfluid mixtures.
We show that in a neutron star at hydrostatic
and beta-equilibrium
the red-shifted temperature gradient
is smoothed out by neutron superfluidity
(but not by proton superfluidity).
We calculate radial oscillation modes of neutron stars
assuming ``frozen''
nuclear composition in the pulsating matter.
The resulting  pulsation frequencies show a strong temperature dependence
in the temperature range $(0.1-1) T_{c{\rm n}}$, where
$T_{c{\rm n}}$ is the critical temperature
of neutron superfluidity.
Combining our results with thermal evolution,
we obtain a significant evolution of the pulsation spectrum, 
associated with highly efficient
Cooper pairing neutrino emission, 
for 20 years after superfluidity onset.
\end{abstract}

\begin{keywords}
stars: neutron -- oscillations -- superfluidity.
\end{keywords}

\section{Introduction}
\label{introduction}
It is commonly accepted that a neutron star 
becomes superfluid (superconducting) at a 
certain stage of its thermal evolution 
(see, e.g., Lombardo and Schulze 2001). 
It is believed, in particular, that protons 
pair in the spin singlet ($^1S_0$) state, while neutrons 
pair in the spin triplet ($^3P_2$) state in the neutron star core.
A large number of different models of nucleon pairing have been
proposed in literature (references to original papers can be found
in Yakovlev et al. 1999 and in Lombardo and Schulze 2001).
These models predict very different density profiles 
of neutron (n) and proton (p) critical 
temperatures, $T_{c{\rm n}}(\rho)$ 
and $T_{c{\rm p}}(\rho)$, respectively.

In spite of the many theoretical uncertainties in the theory, 
it is clear that superfluidity strongly affects 
the neutron star evolution, for example, its cooling
(see, e.g., Yakovlev and Pethick 2004, Page et al. 2004),
neutron star pulsations (see e.g., Mendell 1991a,b, 
Lindblom and Mendell 1994, Lee 1995, Andersson and Comer 2001a, 
Andersson et al. 2002, Prix et al. 2004), 
and is probably related to pulsar glitches 
(see Alpar et al. 1984, Andersson et al. 2003,
Mastrano and Melatos 2005, Peralta et al. 2005).

In this paper we discuss the effect of superfluidity
on neutron star dynamics. 
The hydrodynamics of a superfluid liquid, 
composed of identical particles, 
was formulated by Khalatnikov (1952) 
within Tisza's (1938) two-fluid model, 
which was elaborated by Landau (1941, 1947). 
This ``orthodox'' two-fluid model is based on the 
assumption of two independent velocity fields: 
the ``normal" velocity of thermal excitations ${\pmb V}_{\rm q}$
and the ``superfluid" velocity ${\pmb V}_{\rm s}$, 
each carrying some part of the mass of liquid, 
so that the mass current density ${\pmb j}$ can be written as
\begin{equation}
{\pmb j}=(\rho - \rho_{\rm s}) \, {\pmb V}_{\rm q} 
+ \rho_{\rm s} \, {\pmb V}_{\rm s},
\label{masscurrent}
\end{equation}
where $\rho_{\rm s}$ is known as the superfluid density. 
The superfluid component moves 
without friction and does not interact 
with the normal fluid.
The hydrodynamic equations in this case include the equation of motion 
for the superfluid component, in addition to the energy 
and momentum conservation laws and the 
continuity equations for mass density and entropy 
(see, e.g., Putterman 1974, Landau and Lifshitz 1987, Khalatnikov 1989).

Obviously, the hydrodynamics described above
cannot be applied directly 
to superfluid neutron stars. 
The stellar core consists of, at least, three kinds of particles 
(neutrons, protons, and electrons), 
and neutrons and protons may be superfluid. 
The superfluid hydrodynamics was extended to superfluid mixtures 
by Arkhipov and Khalatnikov (1957) and Khalatnikov (1973) 
and later, more accurately, by Andreev and Bashkin (1975).

The main element of hydrodynamics and kinetics 
of superfluid mixtures is the 
entrainment matrix $\rho_{ik}$, 
which naturally appears in the theory as a generalization
of the superfluid density $\rho_{\rm s}$
to the case of superfluid mixtures. 
If the only baryons in the core are neutrons and protons, 
the matrix $\rho_{ik}$ can be found from the relations 
(Andreev and Bashkin 1975):
\begin{eqnarray}
{\pmb j}_{\rm n} &=& (\rho_{\rm n} - \rho_{\rm nn}-
\rho_{\rm np}) \, {\pmb V}_{\rm q} + 
\rho_{\rm nn} \, {\rm {\pmb V}}_{{\rm n} {\rm s}} +
\rho_{\rm np} \, {\rm {\pmb V}}_{{\rm p} {\rm s}} \,, 
\label{Jn} \\
{\pmb j}_{\rm p} &=& 
(\rho_{\rm p} - \rho_{\rm pp}-
\rho_{\rm pn}) \, {\pmb V}_{\rm q} +
\rho_{\rm pp} \, {\rm {\pmb V}}_{{\rm p} {\rm s}} +
\rho_{\rm pn} \, {\rm {\pmb V}}_{{\rm n} {\rm s}} \,. 
\label{Jp} 
\end{eqnarray}
Here $\rho_i = m_{i} n_{i}$, $m_{i}$ 
is the mass of a free particle and $n_{i}$ is the number density of  
particle species $i$ with $i =$ n or p; 
${\pmb j}_{i}$ and ${\pmb V}_{i {\rm s}}$ 
are the mass current density and the superfluid 
velocity of particle species $i$, 
respectively. 
Since ``normal'' protons and neutrons 
will be locked together by friction, 
we assume that their
velocities ${\pmb V}_{\rm q}$ are identical. 
In other words we assume that
the characteristic time $\tau_{\rm np}$ of neutron-proton collisions
is negligible in comparison with the typical hydrodynamic time 
(e.g.,  the inverse frequency $\omega^{-1}$ of stellar pulsations).
For example, for non-superfluid matter 
$\tau_{\rm np} \sim (10^{-18}-10^{-19}) \, T_9^{-2}$~s 
(see, e.g., Yakovlev and Shalybkov 1991) 
is much smaller than $\omega^{-1} \sim 10^{-4}$~s (see Section~6),
where $T_9=T/(10^9 {\rm K})$ is the temperature
in units of $10^9$ K. 

It follows from the phenomenological analysis 
of Andreev and Bashkin (1975)
that the matrix $\rho_{ik}$ is 
symmetric: $\rho_{\rm np}=\rho_{\rm pn}$. 
Moreover, at zero temperature the equalities
\begin{equation}
\rho_{\rm nn} +\rho_{\rm np} = \rho_{\rm n}, \quad
\rho_{\rm pp} +\rho_{\rm pn} = \rho_{\rm p}
\label{sflT0}
\end{equation}
must hold (see, e.g., Borumand et al. 1996), in order for the system to 
be invariant under Galilean transformations.

The entrainment matrix $\rho_{ik}$ 
for a non-relativistic 
neutron-proton mixture was calculated by Borumand et al. (1996) 
at zero temperature and
by Gusakov and Haensel (2005) for any temperature. 
At $T=0$ 
entrainment coefficients analogous to the matrix $\rho_{ik}$
have also been calculated by Comer and Joynt (2003). 
Even though neutrons (and certainly protons) can be 
considered non-relativistic 
with good accuracy up to the densities $\rho \la 10^{15}$ g cm$^{-3}$,
a fully relativistic calculation of Comer and Joynt (2003) 
is more self-consistent. 
Nevertheless, we will use the results 
obtained by Gusakov and Haensel (2005), because we deal 
with dynamic effects associated with 
finite temperatures in the neutron star core.

The hydrodynamics of superfluid mixtures presented 
by Andreev and Bashkin (1975) cannot be 
applied directly to neutron stars, since 
it is an essentially non-relativistic theory. 
We need to generalize the description to take into account 
the effects of General Relativity
which are important to neutron stars. 
Landau's two-fluid model, initially applied to liquid helium II, was extended to 
General Relativity by Carter (1976, 1979, 1985) 
using a convective variational principle 
and by Khalatnikov and Lebedev (1982) and Lebedev and Khalatnikov (1982) 
on the basis of a potential variational principle. 
The equivalence of these two approaches 
in the non-dissipative limit has been demonstrated 
by Carter and Khalatnikov (1992a,b). 
The former approach was extended by Carter 
and collaborators to analyze
superfluid mixtures, in particular, 
neutron star matter 
(see, e.g., Langlois et al. 1998, Carter and Langlois 1998).
The hydrodynamic equations derived from the convective 
variational principle impose restrictions 
on the ``canonical" coordinates and momenta. 
They include various phenomenological coefficients, 
which need to be related to parameters that are 
actually calculated from microscopic theory, 
for example, the superfluid densities. 
For this reason we will not use Carter's elegant framework here, even though
all available calculations of superfluid oscillations 
in General Relativity have so far been 
made within this approach
(see Comer et al. 1999, Andersson and Comer 2001b, 
Andersson et al. 2002, Yoshida and Lee 2003). 
Instead, we will employ a version of 
superfluid hydrodynamics derived by Son (2001) 
from microscopic theory 
(see also Pujol and Davesne 2003, Zhang 2003). 
Slightly modified, this approach has 
the advantage of offering an easy interpretation
of the various physical quantities entering the hydrodynamic equations. 
Although one can show that our equations are formally equivalent to those 
of Carter, it is clear that further work is needed to connect
his formulation with the microphysics.

The aim of the present study is to analyze the effect
of finite temperatures on pulsations 
of superfluid neutron stars. 
Pulsations may be excited during the star's formation 
or during its evolution under
the action of external perturbations 
(e.g., accretion, gravitational perturbations)
or internal instabilities (associated with unstable pulsation modes).
A possible signature of these pulsations would be the modulation of the electromagnetic
radiation from the neutron star surface or the detection (in the future) 
of gravitational radiation generated by nonaxisymmetric fluid motion.
%
It will be shown that the effect of finite temperatures may
essentially influence the pulsation spectrum in the temperature range
$T \sim (0.1-1) T_{c{\rm n}}$, 
because in this range the entrainment matrix $\rho_{ik}$ 
changes considerably and cannot be treated as a
constant. This, in turn, affects the hydrodynamic equations 
for superfluid mixtures and, hence, 
the oscillations of the star. 
To simplify the problem, we restrict ourselves to the case of
radial pulsations and examine a simple one-fluid model 
of the non-elastic neutron star crust consisting of 
normal matter. The core will be assumed to consist of neutrons, 
protons and electrons (npe-matter), with both 
types of nucleons being superfluid. 

We would like to note that {\it all} 
previous calculations of global pulsations 
of superfluid neutron stars were made 
in a zero-temperature approximation. 
We believe that this is too idealized for two reasons. 
First, even an initially cold star can be heated 
by pulsations because of the transformation of 
the pulsation energy into heat 
(for example, due to viscous dissipation, 
see Gusakov et al. 2005). 
Second, the critical temperatures of nucleons depend on the density. 
This is a bell-shaped curve, which shows 
that the critical temperature first rises with the density 
and then decreases after reaching a 
maximum. Thus, for any given temperature $T$ 
there is usually a region in the star with 
$T \sim T_{c{\rm n}}$. This is an important point that is 
worth emphasizing. 

This paper is organized as follows. 
In Section 2 we extend Son's equations to superfluid 
mixtures and rewrite them using more appropriate variables.
In Section 3 we consider equilibrium configurations of neutron stars. 
%
%
%
In Section 4 we discuss equations for radial pulsations 
taking into account a finite temperature in the core. 
In Section 5 we analyze short wavelength solutions to these equations, 
i.e., sound waves in the superfluid neutron star. 
In Section 6 we examine the numerical solutions 
to the pulsation equations and the eigenfrequency spectrum 
as a function of temperature.
In addition, we study the evolution 
of the oscillation spectrum 
during the star cooling.

\section{
Relativistic equations for non-dissipative 
hydrodynamics of superfluid mixtures}
\label{hydrodynamics_equations}
In this section, the relativistic equations suggested by Son (2001) 
for a one-component superfluid liquid 
at finite temperature 
will be extended to multicomponent mixtures, 
and rewritten in a form which is better suited for our application. 
For simplicity, let us consider a mixture of three kinds of 
particles, assuming that two kinds are superfluid 
and one kind is normal. 
In a neutron star, for example, neutrons 
and/or protons may  be superfluid, 
while electrons (with species index ${\rm e}$) are normal.

It is well known that, in superfluid 
matter, several independent motions 
with different velocities may coexist without dissipation
(see, e.g., Khalatnikov 1989). 
When a mixture is composed 
of two superfluids and one normal fluid
(in principle, there may be many normal species), 
the system is fully defined by three 4-velocities 
$u^{\mu}$, $w_{({\rm n})}^{\mu}$, and $w_{({\rm p})}^{\mu}$.
The latter two arise from 
additional degrees of freedom associated 
with superfluidity. The velocity $u^{\mu}$
refers to electrons as well as ``normal'' neutrons and 
protons (Bogoliubov excitations of neutrons and protons).

If there are several independent motions, 
the question arises how to define the comoving frame
in order to determine
the basic thermodynamic quantities: the energy 
density $\varepsilon$ and the particle number densities 
$n_l$ ($l={\rm n,p,e}$).
Without any loss of generality, we 
can assume that the reference frame in which the velocity $u^{\mu}$
equals to $u^{\mu}= (1,0,0,0)$ is comoving. 
This assumption imposes certain restrictions on 
the particle 4-current $j^{\mu}_{(l)}$ and the energy-momentum 
tensor $T^{\mu \nu}$
\begin{equation}
u_{\mu} j^{\mu}_{(l)} = -n_l, \,\,\,\,\,\,\, 
u_{\mu} u_{\nu} T^{\mu \nu} = \varepsilon.
\label{relation}
\end{equation}
The full set of hydrodynamic equations for superfluid mixtures 
which satisfy these conditions is
\begin{eqnarray}
\dd \varepsilon &=& T \, \dd S + \mu_i \, \dd n_i + \mu_{\rm e} \, \dd n_{\rm e} 
+ { Y_{ik} \over 2} \, \dd \left( w^{\alpha}_{(i)} w_{(k) \alpha} \right),
\label{2ndlaw} \\
j^{\mu}_{ (l) ; \, \mu} &=& 0, \quad 
j^{\mu}_{(i)} = n_i u^{\mu} + Y_{ik} w^{\mu}_{(k)}, \quad
j^{\mu}_{({\rm e})} = n_{\rm e} u^{\mu},
\label{currents} \\
T^{\mu \nu}_{; \,\mu} &=& 0, \quad  
T^{\mu \nu} = (P+\varepsilon) \, u^{\mu} u^{\nu} + P g^{\mu \nu} 
+ Y_{ik} \left( w^{\mu}_{(i)} w^{\nu}_{(k)} + \mu_i \, w^{\mu}_{(k)} u^{\nu} 
+ \mu_k \, w^{\nu}_{(i)} u^{\mu} \right), \,\,\,\,\,
\label{Tmunu} \\
u_{\mu} \, w^{\mu}_{(i)} &=& 0. \,\,\,
\label{sfl} 
\end{eqnarray} 
Here and below, the subscripts $i$ and $k$ 
refer to nucleons: $i,k={\rm n, p}$.
Unless otherwise stated, 
a summation is assumed over repeated spacetime indices (Greek letters)
$\mu$, $\nu$, $\alpha$ and nucleon species indices (Latin letters) $i$, $k$. 
Eq. (\ref{2ndlaw}) represents the second 
law of thermodynamics for superfluid mixtures, 
while Eqs. (\ref{currents}) and (\ref{Tmunu}) describe particle
and energy-momentum conservation laws, respectively. 
Finally, Eq. (\ref{sfl}) 
is the additional equation for a superfluid component;
it is a necessary condition for Eq. (\ref{relation}) to hold.

In Eqs. (\ref{2ndlaw})--(\ref{sfl}) 
$g^{\mu \nu}$ is the metric tensor; 
$S$ is the entropy per unit volume; 
$\mu_l$ is the relativistic chemical potential 
of particle species $l={\rm n,p,e}$; 
$P$ is the pressure which is defined in the same way as for ordinary
(non-superfluid) matter:
\begin{equation}
P= -\varepsilon + \mu_i n_i + \mu_{\rm e} n_{\rm e} + T S.
\label{Pres}
\end{equation}
Finally, $Y_{ik}=Y_{ki}$ 
is a $2 \times 2$ symmetric matrix, 
whose elements are the functions of temperature 
$T$ and the number densities of neutrons and protons. 
Using Eqs. (\ref{2ndlaw}) and (\ref{Pres}), we can write the 
Gibbs-Duhem relation for a superfluid mixture:
\begin{equation}
\dd P = S \, \dd T + n_{i} \, \dd \mu_{i} + n_{{\rm e}} \, \dd \mu_{\rm e} 
- {Y_{ik} \over 2} \, \dd \left( w^{\alpha}_{(i)} w_{(k) \alpha} \right).
\label{Gibbs}
\end{equation}
%

The requirement of constant total entropy of the mixture
imposes an additional constraint on the 
4-velocities $w_{(i)}^{\mu}$.
Namely, we obtain the correct hydrodynamic equations 
for a perfect superfluid mixture 
if the 4-velocities $w_{(i)}^{\mu}$ have the form
\begin{equation}
w^{\mu}_{(i)} = {\partial \phi_i \over \partial x_{\mu}} 
- q_i A^{\mu} - \mu_i u^{\mu},
\label{def}
\end{equation}
where $\phi_i$ is an arbitrary scalar function, 
$A^{\mu}$ is the 4-potential of the electromagnetic field, 
and $q_i$ is the electric charge of nucleon species $i$. 
It is easy to demonstrate that with the quantity $w^{\mu}_{(i)}$ given by
Eq. (\ref{def}), the set of equations (\ref{2ndlaw})--(\ref{sfl}) 
leads to entropy conservation: 
\begin{equation}
(S u^{\mu})_{; \mu} = 0.
\label{entropy}
\end{equation}
Obviously, the entropy is carried
with the same velocity $u^{\mu}$ as the normal fluid, 
i.e., the entropy of the superfluid fraction
in the mixture is zero.

Let us now specify the physical meaning of the quantities 
$\phi_i$, $u^{\mu}$, and $Y_{ik}$.
For this aim we 
will examine how they are related in the non-relativistic limit
to the superfluid velocity ${\pmb V}_{i{\rm s}}$, 
the normal velocity ${\pmb V}_{\rm q}$, 
the wave function phase of 
the 
Cooper-pair
condensate $\Phi_i$, 
and the entrainment matrix $\rho_{ik}$ 
(these quantities appear in the non-relativistic hydrodynamics of 
superfluid mixtures discussed in detail by Andreev and Bashkin 1975). 
One can demonstrate that the following relations hold:
\begin{eqnarray}
{\pmb V}_{i{\rm s}} &=& {1 \over m_i} \left( {\pmb \triangledown} \phi_i 
- q_i {\bf A} \right), 
\quad \quad  
{\pmb \triangledown} \phi_i = {\hbar \over 2} \, {\pmb \triangledown} \Phi_i,
\label{link} \\
{\pmb V}_{\rm q} &=& {\pmb u}, \quad \quad 
Y_{ik} = Y_{ki} = {\rho_{ik} \over m_i m_k}.
\label{matrix}
\end{eqnarray}
Here and below, the speed of light is assumed to be $c=1$.
For convenience, a brief glossary of symbols
is presented in  Table 1.

Let us discuss in more detail the properties of the matrix $Y_{ik}$. 
In the absence of superfluidity, 
when the temperature $T$ is higher than 
the critical temperatures of neutrons $T_{c {\rm n}}$ and 
protons $T_{c {\rm p}}$, we have $Y_{ik}=0$. 
Then the expressions for the 4-currents (\ref{currents}) 
and for the energy-momentum tensor (\ref{Tmunu})
take the standard form and describe a normal perfect fluid
(see, e.g., Landau and Lifshitz 1987). 
If, for example, the inequality $T_{c{\rm n}} < T < T_{c{\rm p}}$ 
holds, i.e., if we have only superfluid protons, 
the only non-vanishing matrix element is $Y_{\rm pp}$. 
In contrast, at $T = 0$ 
{\it all} neutrons and protons form Cooper pairs. 
In other words, there are no nucleons 
moving with the normal fluid component at velocity $u^{\mu}$. 
A 4-current $j^{\mu}_{(i)}$, 
therefore, is independent of $u^{\mu}$, and we have the condition 
(see Eqs. \ref{currents} and \ref{def}):
\begin{equation}
\mu_k \, Y_{ik}(T=0) = n_i.
\label{Tzero}
\end{equation}
Unfortunately, to our best knowledge, results for  
the matrix $Y_{ik}$ at finite temperatures
have not yet been presented in the literature.
Nevertheless, Gusakov and Haensel (2005) 
calculated the entrainment matrix $\rho_{ik}(T)$. 
As we have already mentioned, the matrices $Y_{ik}$ and $\rho_{ik}$ 
are interrelated by Eq. (\ref{matrix}) 
in the non-relativistic limit. 
Thus, we will  use an approximate expression for the matrix $Y_{ik}$, 
which satisfies Eq. (\ref{matrix}) in the non-relativistic limit and 
at the same time meets the condition (\ref{Tzero}):
\begin{equation}
Y_{\rm np} = Y_{\rm pn} = {\rho_{\rm np} \over m_{\rm n} m_{\rm p}}, 
\quad
Y_{\rm nn} = {\rho_{{\rm nn}} +\rho_{\rm np} -m_{\rm n} \mu_{\rm p} Y_{\rm np} 
\over m_{\rm n} \mu_{\rm n}}, 
\quad
Y_{\rm pp} = {\rho_{{\rm pp}} +\rho_{\rm pn} -m_{\rm p} \mu_{\rm n} Y_{\rm pn} 
\over m_{\rm p} \mu_{\rm p}}. 
\label{Yik}
\end{equation}
The condition (\ref{Tzero}) can be derived 
from these formulas, if we take into account that
Eqs. (\ref{sflT0}) must hold at $T = 0$.
\begin{table}
\caption{A brief glossary of symbols describing 
superfluid hydrodynamics in both the non-relativistic and the relativistic domains.
Subscripts $i$ and $k$ refer to nucleons: $i, k$=n, p.}
\begin{tabular}{|l|l|}
\hline
$T_c$               &   critical temperature \\
\hline
$\rho$              &   density              \\
\hline
$\rho_{\rm s}$      &   superfluid density   \\
\hline
${\pmb V_{\rm q}}$  &   velocity of thermal excitations (Bogoliubov quasiparticles) \\
\hline
${\pmb V_{\rm s}}$  &   superfluid velocity \\
\hline
${\pmb j}$          &   mass current density \\ 
\hline
\hline
$T_{ci}$           &    critical temperature of particles $i$ \\
\hline
$\rho_{i}$         &    density of particles $i$ \\
\hline
$\rho_{ik}$        &    entrainment matrix \\
\hline
$\Phi_i$           &    wave function phase of the Cooper-pair condensate of particles $i$\\
\hline
${\pmb V_{i{\rm s}}}$ & superfluid velocity of particles $i$ \\
\hline
${\pmb j}_i$          & mass current density of particles $i$ \\
\hline
\hline
$n_i$                &   number density of particles $i$ \\
\hline
$Y_{ik}$             &   relativistic entrainment matrix, 
                         in the non-relativistic limit $Y_{ik}=\rho_{ik}/m_{i} m_{k}$ \\
\hline
$u^{\mu}$            &   4-velocity of electrons and neutron and proton thermal excitations \\
\hline
$\phi_i$             &   scalar potential related to $\Phi_i$ by 
                         ${\pmb \triangledown} \phi_i = \hbar \,\,{\pmb \triangledown} \Phi_i/2$ \\
\hline
$w_{(i)}^{\mu}$      &   4-velocity which reduces to
                         ${\pmb w}_{(i)}=m_i ({\pmb V}_{i{\rm s}}-{\pmb V}_{\rm q})$ 
			 in the non-relativistic limit\\
\hline
$j^{\mu}_{(i)}$      &   4-current of particles $i$ \\
\hline
$A^{\mu}$            &   4-potential of the electromagnetic field \\
\hline
$q_i$                &   electric charge of particles $i$\\
\hline
\hline
\end{tabular}
\end{table}

\section{Equilibrium configurations of superfluid neutron stars}
Let us now use the above formulas to describe neutron stars. 
For simplicity, consider a non-rotating star. 
We will often refer to the results of the pioneering work of Chandrasekhar (1964)
devoted to radial pulsations of non-superfluid stars in General Relativity. 
The metric for a spherically symmetric star, 
which experiences radial pulsations, can be written as 
(see, e.g., Chandrasekhar 1964)  
\begin{equation}
 \dd s^2 =  -{\rm e}^{\nu} \dd t^2 + r^2 \dd \Omega^2
    + {\rm e}^{\lambda} \, \dd r^2,
\label{ds}
\end{equation}
where $r$ and $t$ are the radial and time coordinates, 
respectively; 
$\dd \Omega$ is a solid angle element 
in a spherical frame with the origin at 
the stellar center.
The metric functions $\nu$ and $\lambda$
depend only on $r$ and $t$. 
The quantities referring to 
a star in hydrostatic equilibrium
will be marked with the subscript ``0";
in particular, the metric coefficients 
of an unperturbed star will be 
denoted as $\nu_0(r)$ and $\lambda_0(r)$.

In the equilibrium neutron star the measurable physical 
quantities (e.g., the number densities) must be time-independent.
Thus, the continuity equation for electrons 
(\ref{currents})  and the 
expression for the 4-velocity of the normal component
\begin{equation}
u^{\mu} = {\dd x^{\mu} \over \dd s}
\label{umu}
\end{equation}
yield (for a spherically symmetric star!)
\begin{equation}
u^{0} = {\rm e}^{-\nu_0/2}, \quad u^1=u^2=u^3=0.
\label{umustat}
\end{equation}
Next, the continuity equations for neutrons and protons (\ref{currents}) give
\begin{equation}
w^{1}_{(i)}=w^{2}_{(i)}=w^{3}_{(i)}=0.
\label{wstat}
\end{equation}
Finally, in view of Eq. (\ref{umustat}), 
one obtains from Eq. (\ref{sfl})
\begin{equation}
w^{0}_{(i)} = 0.
\label{w0}
\end{equation}
It is clear from Eqs. (\ref{umustat})--(\ref{w0}) 
that the energy-momentum tensor (\ref{Tmunu}) of an 
equilibrium superfluid star is the same as that 
of a non-superfluid one. 
Therefore, the formulas that describe 
hydrostatic equilibrium of non-superfluid stars 
can be applied to our case as well. 
In particular, the following formula is valid 
(see, e.g., equation 21 of Chandrasekhar 1964)
\begin{equation}
{\dd P_0 \over \dd r} = - {1 \over 2} \, \left(P_0 + \varepsilon_0 \right) \,
{\dd \nu_0 \over \dd r}.
\label{hydro}
\end{equation}
New information can be obtained from Eq. (\ref{w0}). 
When written for neutrons, it gives, together with Eq. (\ref{def}),
\begin{equation}
{ \partial \phi_{\rm n0} \over \partial t} = - \mu_{{\rm n} 0} \, {\rm 
e}^{\nu_0/2}.
\label{statcond}
\end{equation}
On the other hand, from Eqs. (\ref{def}), (\ref{umustat}) 
and ({\ref{wstat}}) one finds 
\begin{equation}
{ \partial \phi_{\rm n0} \over \partial r} = 0.
\label{dphidr}
\end{equation}
It follows from Eqs. (\ref{statcond}) and (\ref{dphidr}) that
\begin{equation}
{\dd \over \dd r} \left( \mu_{{\rm n} 0} \, {\rm e}^{\nu_0/2} \right) =0.
\label{muncond}
\end{equation}
It should be emphasized that the application of conditions (21) and (22) to 
protons will not yield a constraint 
similar to Eq. (\ref{muncond}) for $\mu_{{\rm p}0}$, because Eq. (\ref{def}) for 
the protons depends, additionally, on the 4-potential of the electromagnetic field. 
We are not interested here in 
the relation between $A^{\mu}$ and $\mu_{{\rm p}0}$ that 
can be derived from Eqs. (\ref{wstat}) and (\ref{w0}). 

Assuming that a star at hydrostatic equilibrium meets, in addition, the 
quasineutrality condition, $n_{{\rm e}0} = n_{{\rm p}0}$,
%
%
one gets from Eqs. (\ref{Pres}) and (\ref{Gibbs})
\begin{eqnarray}
P_0 + \varepsilon_0 &=& \mu_{{\rm n}0} n_{{\rm b}0} + \delta \mu_0 n_{\rm e0} + T_0 S_0,
\label{eq1} \\
{\dd P_{0} \over \dd r} &=& n_{{\rm b}0} \, {\dd \mu_{{\rm n}0} \over \dd r }
+n_{{\rm e}0} \, {\dd \delta \mu_{0} \over \dd r } 
+ S_0 \, {\dd T_0 \over \dd r},
\label{eq2} 
\end{eqnarray}
where $n_{\rm b0} \equiv n_{\rm n0}+n_{\rm p0}$ is the baryon number density;
$\delta \mu_0 \equiv \mu_{\rm p0}+\mu_{\rm e0}-\mu_{\rm n0}$. 
Substituting the expression for 
$\dd \nu_0/\dd r$ from Eq. (\ref{muncond}) 
into Eq. (\ref{hydro}) and using Eq. (\ref{eq1}), one obtains
\begin{equation}
{\dd P_{0} \over \dd r} = n_{{\rm b}0} \, {\dd \mu_{{\rm n}0} \over \dd r }
-{1 \over 2} \, \left( \delta \mu_0 n_{\rm e0} + T_0 S_0 \right)\, 
{\dd \nu_0 \over \dd r}.
\label{eq22}
\end{equation}
A comparison of Eqs. (\ref{eq2}) and (\ref{eq22}) leads to the equality
\begin{equation}
n_{\rm e0} \, {\dd \over \dd r} \left(\delta \mu_0 {\rm e}^{\nu_0/2} \right)
+ S_0 \, {\dd \over \dd r} \left(T_0 {\rm e}^{\nu_0/2} \right)=0.
\label{gradT1}
\end{equation}
Note that, to derive this formula we have considered
a star in hydrostatic equilibrium
(but not necessarily in thermal, diffusive, 
or beta-equilibrium). 
If we assume, in addition,
that in some region of the star
($i$) the thermal equilibrium condition is fulfilled
\begin{equation}
{\dd \over \dd r} \left(T_0 {\rm e}^{\nu_0/2} \right)=0,
\label{gradT}
\end{equation}
and ($ii$) neutrons are superfluid, 
then Eq. (\ref{gradT1}) tells us that this region must be
in diffusive equilibrium 
(the opposite statement is also correct:
diffusive equilibrium means thermal equilibrium 
for the problem in question).
Indeed, in this case we have 
from Eqs. (\ref{muncond}) and (\ref{gradT1})
\begin{equation}
{\dd \over \dd r} \left( \mu_{{\rm n} 0} \, {\rm e}^{\nu_0/2} \right)=0, \quad
{\dd \over \dd r} \left[ \left(\mu_{{\rm p} 0}+ \mu_{{\rm e} 0} \right)\, 
{\rm e}^{\nu_0/2} \right] =0.
\label{gradT2}
\end{equation}
These conditions describe the diffusive equilibrium 
of npe-matter and are quite standard 
(see, e.g., Landau and Lifshitz 1980). 
The second condition of Eq. (\ref{gradT2}) is nothing but a sum of 
the diffusive equilibrium conditions written
for protons and electrons.
Each of them includes 
a self-consistent electrostatic potential
to ensure quasineutrality 
(the most recent discussion 
of diffusive equilibrium as applied to npe-matter of neutron stars 
is given by Reisenegger et al. 2006). 
We are not interested here in determining
this potential: 
it cancels out after the summation.

In this paper we assume that an unperturbed star is at
hydrostatic and beta-equilibrium (i.e. $\delta \mu_0=0$).
In this special case one immediately 
obtains from Eq. (\ref{gradT1}) 
the thermal equilibrium condition (\ref{gradT}).
Thus, we arrive at the conclusion 
that a (red-shifted) temperature gradient 
cannot exist in any region of 
a hydrostatically and beta-equilibrated
neutron star 
which contains superfluid neutrons. 
This situation is identical to that for pure helium II 
(see, e.g., Khalatnikov 1989). 
Note that, proton superfluidity 
imposes no such restrictions on the temperature gradient.

\section{Radial pulsations of superfluid neutron stars}
In this section we consider a star 
with small radial perturbations. 
Accordingly, in all the equations 
we shall neglect the quantities which
are second order and higher 
in the pulsation amplitude and retain the linear terms.
In addition, we will use the hypothesis of a frozen nuclear
composition, neglecting the effect of 
beta-processes on the chemical composition 
of the core during the pulsations. 
This assumption is justified if 
the radial pulsation frequencies 
are $\omega \gg 1/\tau$, where 
$\tau$ is the characteristic time 
of beta-equilibration. 
Recall that for non-superfluid matter and under 
the condition $|\mu_{\rm p}+\mu_{\rm e}-\mu_{\rm n}| \ll T$
it can be estimated that 
$\tau \sim T_9^{-6}$ 
months
if beta-relaxation proceeds via the modified Urca process
(see, e.g., Yakovlev et al. 2001);
for superfluid matter beta-relaxation rates were calculated
by Haensel et al. (2000, 2001), 
and by Villain and Haensel (2005).
The final assumption we make is the 
validity of the quasineutrality condition 
in a pulsating star,
\begin{equation}
n_{\rm e}=n_{\rm p},
\label{quasineutrality}
\end{equation}
which should hold
since $\omega$ is much smaller than 
the plasma frequency of electrons, $\omega_{\rm pe}$. 
In the following, the quantities containing 
no ``0" subscript refer to a perturbed star. 
If $A$ is a 
physical quantity in a perturbed star and $A_0$ the same quantity  
in the unperturbed star, 
then we denote $A-A_0 \equiv \delta A$. 

The quasineutrality condition leads to 
equal 4-currents of electrons and protons:
\begin{equation}
j^{\mu}_{({\rm e})}=j^{\mu}_{({\rm p})}.
\label{quasineutrality1}
\end{equation}
By substituting the expressions for the currents 
from Eq. (\ref{currents}), one gets 
\begin{equation}
Y_{{\rm p}k} w^{\mu}_{(k)} = 0.
\label{quasineutrality2}
\end{equation}
We will also need the continuity equation for baryons, 
which can be found by summing the 
continuity equations (\ref{currents}) 
for protons and neutrons. With Eq. (\ref{quasineutrality2}), we obtain
\begin{equation}
\left(n_{\rm b} u^{\mu} + Y_{{\rm n}k} w^{\mu}_{(k)} \right)_{; \mu} =0.
\label{baryons}
\end{equation}
%

\subsection{Basic equations}
Using the metric (\ref{ds}), 
one can write the linearized 4-velocity $u^{\mu}$ as
%
\begin{equation}
u^0 = {\rm e}^{-\nu/2},  \quad \quad u^1={\rm e}^{-\nu_0/2} \, v, \quad\quad 
u^2=u^3=0,
\label{umu1}
\end{equation}
where $v \equiv dr/dt$ is the velocity of the normal component 
of the mixture in the radial direction. 
(Note a misprint in formula 25
of Chandrasekhar 1964: 
the expressions for $u^0$ and $u_0$ must 
have $\nu$ instead of $\nu_0$.) 
Using Eq. (\ref{umu1}), one can find 
directly from Eq. (\ref{sfl})
\begin{equation}
w^0_{(i)}=0.
\label{weq0}
\end{equation}
In addition, because particles move
only in the radial direction, we have
\begin{equation}
w^2_{(i)}=w^3_{(i)}=0.
\label{weq23}
\end{equation}
Therefore, the only non-zero components of 
the energy-momentum tensor are
\begin{eqnarray}
T_0^{\, 0}&=& - \varepsilon,  \quad \quad \quad T_1^{\, 1}=T_2^{\, 2}=T_3^{\, 
3}=P, 
\label{components1} \\
T_0^{\, 1} &=& -(P_0+\varepsilon_0) \, v + \Delta T_0^{\, 1},
\label{components2}\\
T_1^{\, 0} &=& {\rm e}^{\lambda_0-\nu_0} \,(P_0+\varepsilon_0) \, v 
- {\rm e}^{\lambda_0-\nu_0} \, \Delta T_0^{\, 1}.
\label{components3}
\end{eqnarray}
These formulas differ from those for 
the normal liquid only 
by the term $\Delta T_0^{\, 1}$,
which is (see Eq. \ref{Tmunu})
\begin{equation}
\Delta T_0^{\, 1} = \mu_{k0} Y_{ik} \,\, u_0 \, w^1_{(i)} 
= -\mu_{\rm n0} Y_{{\rm n} i} \,\, w^1_{(i)} \,{\rm e}^{\nu_0/2}.
\label{dT1}
\end{equation}
When writing the last equality, 
we have used Eq. (\ref{quasineutrality2}) 
and the expression $u_0=-{\rm e}^{\nu/2}$ 
(Eq. \ref{sfl} yields $w^1_{(i)} \sim v$, so that $\nu$ 
can be substituted for $\nu_0$ in Eq. \ref{dT1}). 
Note an important consequence of Eq. (\ref{dT1}): 
if neutrons in a star are normal ($Y_{{\rm n}i}=0$), 
its pulsations will be {\it indiscernible} from those of 
a common non-superfluid star, no matter whether 
the protons are superfluid or not.

Let us analyze Eq. (\ref{weq0}) for neutrons. 
With Eq. (\ref{def}) it can be rewritten as
\begin{equation}
-{\rm e}^{-\nu} \, {\partial \phi_{\rm n} \over \partial t} 
- \mu_{\rm n} \, {\rm e}^{-\nu/2} = 0.
\label{sfleq1}
\end{equation}
By substituting 
$\nu = \nu_0 + \delta \nu(r,t)$, 
$\phi_{\rm n}= \phi_{\rm n0}+\delta \phi_{\rm n}(r,t)$, 
$\mu_{\rm n}=\mu_{\rm n0}+\delta \mu_{\rm n}(r,t)$
into Eq. (\ref{sfleq1}) and using 
Eq. (\ref{statcond}), 
we get
\begin{equation}
{\partial \delta \phi_{\rm n} \over \partial t} =  
- \left(\delta \mu_{\rm n} \, 
+ {1 \over 2}\,\, \mu_{\rm n0} \, \delta \nu \right) \, {\rm e}^{\nu_0/2}.
\label{sfleq2}
\end{equation}
On the other hand, in the linear approximation 
and in view of Eq. (\ref{dphidr}), we have
\begin{equation}
w^1_{\rm (n)} = {\rm e}^{-\lambda} {\partial \phi_{\rm n}\over \partial r}
-\mu_{\rm n} u^{1}=
{\rm e}^{-\lambda_0} {\partial \delta \phi_{\rm n}\over \partial r}
-\mu_{\rm n0} \, {\rm e}^{-\nu_0/2} \,v.
\label{sfleq3}
\end{equation}
By combining Eqs. (\ref{sfleq2}) and (\ref{sfleq3}), we find
\begin{equation}
{\partial \over \partial t} 
\left( {\rm e}^{\lambda_0} \, w^1_{({\rm n})} 
+ \mu_{\rm n0} \, {\rm e}^{\lambda_0-\nu_0/2} \, v \right) = 
-{\partial \over \partial r} 
\left( \delta \mu_{\rm n} \, {\rm e}^{\nu_0/2} 
+ {1 \over 2} \,\, \mu_{\rm n0} \, {\rm e}^{\nu_0/2}\, \delta \nu \right).
\label{sfleq4}
\end{equation}
Let us introduce new variables 
$z_i$ and $\xi$ according to 
(there is no summation over $i$ here!):
\begin{eqnarray}
w^1_{(i)}&=&\mu_{ i0} \, {\rm e}^{-\nu_0/2} \, {\partial z_i \over \partial t}, 
\label{change1} \\
v &=& {\partial \xi \over \partial t}.
\label{v} 
\end{eqnarray}
The time integration of
Eq. (\ref{quasineutrality2}) gives the relation between 
the variables $z_{\rm n}$ and $z_{\rm p}$ 
\begin{equation}
\mu_{k0} \, Y_{{\rm p}k} \, z_{k} = 0.
\label{quasineutrality3}
\end{equation}
Assuming now that all perturbations 
vary with time as ${\rm exp}(i \omega t)$, 
we rewrite Eq. (\ref{sfleq4}) in the form:
\begin{equation}
\mu_{\rm n0} \,\, {\rm e}^{\lambda_0-\nu_0/2} \,\, \omega^2 \,
\left( z_{\rm n} + \xi \right)
={\partial \over \partial r} 
\left( \delta \mu_{\rm n} \, {\rm e}^{\nu_0/2} 
+ {1 \over 2} \,\, \mu_{\rm n0} \, {\rm e}^{\nu_0/2}\, \delta \nu \right).
\label{sfleq5}
\end{equation}
Thus, we have derived one of the equations 
that describe pulsations of a relativistic superfluid star. 
There is no analogue of this equation for non-superfluid stars. 
In order to 
determine
the unknown 
eigenfunctions $z_i$ and $\xi$ and the frequency spectrum, 
it is necessary to find an additional pulsation equation. 
In principle, this can be done by writing
Einstein's  equations with the 
energy-momentum tensor given by 
Eqs. (\ref{components1})--(\ref{components3}). 
However, the situation can be considerably simplified 
because this energy-momentum tensor does not 
essentially differ from that used by 
Chandrasekhar (1964) in the analysis of pulsations
of non-superfluid stars 
(see his Eqs. 27 and 28). 
By adjusting his derivation to our case, 
we find the following expressions for the 
quantities $\delta \lambda$,
$\delta \varepsilon$, and $\partial \delta \nu/ \partial r$:
\begin{eqnarray}
\delta \lambda &=& 
\widetilde{T}_0^{\, 1} \,\,
{1 \over P_0 + \varepsilon_0} \,
\,{\dd \over \dd r} \left( \lambda_0 + \nu_0 \right), 
\label{dlambda} \\
\delta \varepsilon &=& {1 \over r^2} \, {\partial \over \partial r} 
\left( r^2 \, \widetilde{T}_0^{\, 1}  \right),
\label{depsilon} \\
{\partial \delta \nu \over \partial r} &=& 
{1 \over P_0 + \varepsilon_0} \,
\left[ \delta P + \left( {\dd \nu_0 \over \dd r} +{1 \over r} \right) 
\, \widetilde{T}_0^{\, 1} \right] \,
{\dd \over \dd r} \left( \lambda_0 + \nu_0 \right).
\label{dnu} 
\end{eqnarray}
Here, the quantity $\widetilde{T}_0^{\, 1}$ 
is defined by
\begin{equation}
T_0^{\, 1}= 
{\partial \widetilde{T}_0^{\, 1} \over \partial t}
\label{T01}
\end{equation}
and is found to be 
(see Eqs. \ref{components2}, \ref{dT1}, \ref{change1}, and \ref{v})
\begin{equation}
\widetilde{T}_0^{\, 1} = - \left( P_0 + \varepsilon_0 \right) \xi 
- \mu_{\rm n0} \mu_{i0} Y_{{\rm n}i} \, z_i.
\label{T012}
\end{equation}
Eqs. (\ref{dlambda})--(\ref{dnu}) 
are generalizations of the expressions (36), (37), and (41) 
from the paper by Chandrasekhar (1964). 
The pulsation equation (43) of his work 
can be rewritten in our case as
\begin{equation}
- {\rm e}^{\lambda_0-\nu_0} \, \omega^2 \, \widetilde{T}_0^{\, 1} =
{\partial \delta P \over \partial r} + \delta P \, {\dd \over \dd r} 
\left( {1 \over 2} \lambda_0 + \nu_0 \right) + {1 \over 2} \, \delta \varepsilon 
\, 
{\dd \nu_0 \over \dd r} + {1 \over 2} \, \widetilde{T}_0^{\, 1} \, 
\left( {\dd \nu_0 \over \dd r} +{1 \over r} \right) \, {\dd \over \dd r} 
\left( \lambda_0 + \nu_0 \right).
\label{pulseq}
\end{equation}
Eqs. (\ref{sfleq5}) and (\ref{pulseq}) 
fully describe radial pulsations of superfluid neutron stars. 
What remains to be done is to find the unknown functions
$\delta P$ and $\delta \mu_{\rm n}$ entering these equations.

\subsection{The functions $\delta P$ and $\delta \mu_{\rm n}$}
With the quasineutrality condition, 
which is valid in a pulsating neutron star, 
any thermodynamic function 
(for a stellar core composed of neutrons, 
protons, and electrons) 
can be represented as a function 
of three thermodynamic variables, say, 
$n_{\rm b}$, $n_{\rm e}$, and $S$ 
(the quadratically small dependence of the thermodynamic 
parameters on $w^{\alpha}_{(i)} w_{(k) \alpha}$ is neglected). 
Since the pulsations are assumed to be small, 
the pressure 
$P(n_{\rm b}, n_{\rm e}, S)=P_0 + \delta P$ 
and the neutron chemical potential 
$\mu_{\rm n}(n_{\rm b}, n_{\rm e}, S) 
= \mu_{\rm n0}+ \delta \mu_{\rm n}$ 
can be expanded in the vicinity 
of their equilibrium values,
\begin{eqnarray}
\delta P &=& {\partial P(n_{\rm b0}, n_{\rm e0}, S_0) \over \partial n_{\rm b0}} 
\, 
\delta n_{\rm b}
+{\partial P(n_{\rm b0}, n_{\rm e0}, S_0) \over \partial n_{\rm e0}} \,
\delta n_{\rm e}
+{\partial P(n_{\rm b0}, n_{\rm e0}, S_0) \over \partial S_{\rm 0}} \,
\delta S,
\label{expand1} \\
\delta \mu_{\rm n} &=& {\partial \mu_{\rm n}(n_{\rm b0}, n_{\rm e0}, S_0) \over 
\partial n_{\rm b0}} \, 
\delta n_{\rm b}
+{\partial \mu_{\rm n}(n_{\rm b0}, n_{\rm e0}, S_0) \over \partial n_{\rm e0}} 
\,
\delta n_{\rm e}
+{\partial \mu_{\rm n}(n_{\rm b0}, n_{\rm e0}, S_0) \over \partial S_{\rm 0}} \,
\delta S.
\label{expand2}
\end{eqnarray}
Let us find 
$\delta n_{\rm b}$, $\delta n_{\rm e}$, and $\delta S$
from the continuity equations 
for baryons (\ref{baryons}), electrons (\ref{currents}), 
and entropy (\ref{entropy}), respectively. 
Writing explicitly the covariant derivative 
in the metric of Eq. (\ref{ds}) 
and keeping only terms linear in the perturbations, 
one can rewrite the continuity equation for 
baryons (\ref{baryons}) as
\begin{eqnarray}
{\rm e}^{-\nu_0/2} \, {\partial \delta n_{\rm b} \over \partial t} 
+ {1 \over r^2} \, {\partial \over \partial r}
\left( r^2 \, n_{\rm b0} \, {\rm e}^{-\nu_0/2} \, v  \right) +
{1 \over 2} \,\,  n_{\rm b0} \, {\rm e}^{-\nu_0/2} \, \,
{\partial \delta \lambda \over \partial t}
+ {1 \over 2} \, \, n_{\rm b0} \,  {\rm e}^{-\nu_0/2} \, \, v \, {\dd \over \dd 
r} 
\left( \lambda_0 + \nu_0 \right)
\nonumber \\
+ {1 \over r^2} \, {\partial \over \partial r} 
\left( r^2 \, Y_{{\rm n}k} w^1_{(k)} \right) 
+ {1 \over 2} \,\, Y_{{\rm n}k} w^1_{(k)} \, {\dd \over \dd r} 
\left( \lambda_0 + \nu_0 \right) =0. \quad 
\label{baryons2}
\end{eqnarray}
The time integration of this equation 
using Eqs. (\ref{change1}), (\ref{v}), (\ref{dlambda}), 
and the equality (\ref{eq1}) with $\delta \mu_0=0$ will yield
\begin{equation}
\delta n_{\rm b} = - {{\rm e}^{\nu_0/2} \over r^2} \, 
{\partial \over \partial r}
\left(  r^2 \,n_{\rm b0} \,  \xi \, {\rm e}^{-\nu_0/2}\right)
-{{\rm e}^{\nu_0/2} \over r^2} \, {\partial \over \partial r}
\left(r^2 \, \mu_{k0} Y_{{\rm n}k} \, z_k \, {\rm e}^{-\nu_0/2} \right).
\label{baryons3}
\end{equation}
The expressions for $\delta n_{\rm e}$ and $\delta S$ 
can be derived in a similar way:
\begin{eqnarray}
\delta n_{\rm e} &=& - {{\rm e}^{\nu_0/2} \over r^2} \, {\partial \over \partial 
r}
\left(  r^2 \,n_{\rm e0} \, \xi \, {\rm e}^{-\nu_0/2} \right),
\label{electrons}\\
\delta S &=& - {{\rm e}^{\nu_0/2} \over r^2} \, {\partial \over \partial r}
\left(  r^2 \,S_0 \, \xi \, {\rm e}^{-\nu_0/2} \right).
\label{entropy2}
\end{eqnarray}
Eqs. (\ref{baryons3})--({\ref{entropy2}}) 
are generalizations of Chandrasekhar's (1964) equation (50). 
Note that, Eq. (\ref{baryons3}) can 
be rewritten in a more compact form. 
By multiplying its left- and right-hand sides by $\mu_{\rm n0}$ 
and using Eqs. (\ref{muncond}), (\ref{eq1}), (\ref{gradT}), 
(\ref{depsilon}), and (\ref{T012}), 
we find
\begin{equation}
\mu_{\rm n0} \, \delta n_{\rm b} = \delta \varepsilon - T_0 \, \delta S.
\label{baryons4}     
\end{equation}
This is just the second law of 
thermodynamics (\ref{2ndlaw}) 
with the quasineutrality ($n_{\rm e0}=n_{\rm p0}$) 
and beta-equilibrium ($\delta \mu_0 =0$) conditions 
valid for an equilibrium star taken into account.

The substitution of Eqs. (\ref{baryons3})--(\ref{entropy2}) 
into (\ref{expand1}) and (\ref{expand2}) gives, after
standard transformations,
\begin{eqnarray}
\delta P &=& - {\dd P_0 \over \dd r} \, \xi 
- \gamma_1 \, P_0 \, \Phi
-\beta_1 \, P_0 \, \Psi,
\label{dp2} \\
\delta \mu_{\rm n} &=& - {\dd \mu_{\rm n0} \over \dd r} \, \xi 
- \gamma_2 \, \mu_{\rm n0} \, \Phi
- \beta_2 \, \mu_{\rm n0} \, \Psi,
\label{dmu2}     
\end{eqnarray}
with
\begin{eqnarray}
\Phi &=& {{\rm e}^{\nu_0/2} \over r^2} \,
{\partial \over \partial r} 
\left( r^2 \, \xi \, {\rm e}^{-\nu_0/2} \right), \quad \quad
\Psi = {{\rm e}^{\nu_0/2} \over n_{\rm b0} r^2} \,
{\partial \over \partial r} 
\left( r^2 \, \mu_{k0} Y_{{\rm n}k} \, z_k \, {\rm e}^{-\nu_0/2} \right),
\label{psi1} \\    
\gamma_1 &=& {n_{\rm b0} \over P_0} \, 
{\partial P(n_{\rm b0}, x_{\rm e0}, x_{\rm s0}) \over \partial n_{\rm b0}},\quad 
\quad
\gamma_2 = {n_{\rm b0} \over \mu_{\rm n0}} \, 
{\partial \mu_{\rm n}(n_{\rm b0}, x_{\rm e0}, x_{\rm s0}) \over \partial n_{\rm 
b0}},
\label{gamma} \\
\beta_1 &=& {n_{\rm b0} \over P_0} \, 
{\partial P(n_{\rm b0}, n_{\rm e0}, S_0) \over \partial n_{\rm b0}},\quad 
\quad
\beta_2 = {n_{\rm b0} \over \mu_{\rm n0}} \, 
{\partial \mu_{\rm n}(n_{\rm b0}, n_{\rm e0}, S_0) 
\over \partial n_{\rm b0}},
\label{beta1} 
\end{eqnarray}
where
$x_{\rm e0} \equiv n_{\rm e0}/n_{\rm b0}$ and 
$x_{\rm s0} \equiv S_0/n_{\rm b0}$.
It should be noted that the partial derivatives in 
Eq. (\ref{gamma}) are taken at 
constant values of $x_{\rm e0}$ and $x_{\rm s0}$. 
The new parameter $\gamma_1$ 
is 
just
an adiabatic index of matter that 
describes pulsations of normal (non-superfluid) stars. 
When calculating the partial derivatives 
of thermodynamic parameters, 
one can neglect the temperature effects and put 
$S_0 = 0$ and $x_{\rm s0}=0$ everywhere.

Thus, we have found the functions 
$\delta \mu_{\rm n}$ and $\delta P$
under the assumption of frozen nuclear composition. 
In this work, all actual calculations 
of the eigenfrequency spectrum
are based on this assumption. 
Still, we would like to make 
a comment on how one could find 
these functions in the opposite  
case when $\omega \ll 1/\tau$ 
(when the core is in beta-equilibrium during pulsations). 
The pressure $P$ and the 
neutron chemical potential $\mu_{\rm n}$ 
are then functions of $n_{\rm b}$ and $S$ only, 
whereas the electron number density 
$n_{\rm e}(n_{\rm b}, S)$ 
is derived from the beta-equilibrium condition. 
Using Eqs. (\ref{baryons3}) and (\ref{entropy2}), 
one can write
%
%
%
\begin{eqnarray}
\delta P &=& {\partial P(n_{\rm b0}, S_0) \over \partial n_{\rm b0}} \, \delta 
n_{\rm b} 
+ {\partial P(n_{\rm b0}, S_0) \over \partial S_{0}} \, \delta S,
\label{fulleq1} \\
\delta \mu_{\rm n} &=& 
{\partial \mu_{\rm n}(n_{\rm b0}, S_0) \over \partial n_{\rm b0}} \, \delta 
n_{\rm b} 
+ {\partial \mu_{\rm n}(n_{\rm b0}, S_0) \over \partial S_{0}} \, \delta S.
\label{fulleq2} 
\end{eqnarray}
If we now neglect the entropy dependence of 
thermodynamic parameters 
(as is justified for 
the frozen nuclear composition), 
we will arrive at a qualitatively wrong result, 
where one of the branches 
of the pulsation spectrum is missing. 
Indeed, the pulsation equation (\ref{pulseq}) 
will then depend only on the 
eigenfunction $\widetilde{T}_0^{\, 1}$  
(see Eqs. \ref{depsilon}, \ref{baryons4}, and \ref{fulleq1}). 
Therefore, the pulsation eigenfrequencies can be found 
just from Eq. (\ref{pulseq}) alone, 
independently of Eq. (\ref{sfleq5}). 
[It will be shown in the next section that the 
boundary conditions for the pulsation equations 
\ref{sfleq5} and \ref{pulseq} 
can also be formulated only in terms  
of the eigenfunction $\widetilde{T}_0^{\, 1}$.] 
The obtained branch of the pulsation spectrum 
practically coincides with the spectrum 
of a non-superfluid star, 
while the specifically ``superfluid"
pulsation modes will 
be lost. 
To the best of our knowledge, 
such ``temperature" pulsation modes 
have not been discussed previously in the neutron-star literature.

\subsection{Boundary conditions}
Pulsation equations (\ref{sfleq5}) and (\ref{pulseq}) 
together with 
Eqs. (\ref{quasineutrality3}), (\ref{depsilon}), (\ref{dnu}),
({\ref{T012}}), (\ref{dp2}), and (\ref{dmu2})
enable one to determine the unknown functions 
$z_{\rm n}$, $z_{\rm p}$, $\xi$ and the frequency spectrum, 
provided that the boundary conditions are known.

To formulate the boundary conditions, 
we should specify the model problem to be solved. 
We assume neutrons to be superfluid 
inside a sphere of circumferential radius $R_{0}$ 
with $R_0 \leq R_{\rm cc}$, where 
$R_{\rm cc}$ is the 
radial coordinate 
of the crust-core interface. 
Outside the sphere, neutrons are assumed to be normal. 
The parameters 
related to
the outer 
($r>R_0$) 
region of the star will be marked 
with the letter ``${\rm c}$". 
On the stellar surface, 
we have a standard boundary condition: 
\begin{equation}
P_{\rm c}(R + \xi_{\rm c}(R)) = 0,
\label{pc}     
\end{equation}
which can be rewritten as
\begin{equation}
\left[ \delta P_{\rm c} + {\dd P_0 \over \dd r} 
\, \xi_{\rm c} \right]_{r=R} = 0.
\label{pc1}     
\end{equation}
Here $R$ is the circumferential radius of an unperturbed star;
$\xi_{\rm c}$ is the Lagrangian displacement 
of 
matter in the outer region.
In Eq. (\ref{pc1}) we defined 
$P_{\rm c}(R) \equiv P_0(R)+\delta P_{\rm c}$.
All derivatives with respect to $r$ 
at the stellar center must be finite, 
which means that the following limits are finite
\begin{equation}
\lim\limits_{r \rightarrow 0} \xi/r < \infty, \quad \quad
\lim\limits_{r \rightarrow 0} z_{i}/r< \infty.
\label{lim}     
\end{equation}
The other boundary conditions 
should be
formulated 
at the superfluid-normal interface. 
First, the electron current at the interface 
must be continuous. 
It follows then from Eq. (\ref{electrons}) 
that the Lagrangian displacement of ``normal" particles is 
continuous, which leads to
\begin{equation}
\xi(R_0) = \xi_{\rm c}(R_0).
\label{boundary1}     
\end{equation}
In addition, the energy and momentum currents
through the interface must also be continuous. 
These conditions lead to the following equalities 
(see Eqs. \ref{components1} -- \ref{components3} 
together with the expressions \ref{v} and \ref{T012}):
\begin{eqnarray}
P(R_0+\xi(R_0)) &=& P_{\rm c}(R_0+\xi_{\rm c}(R_0)),
\label{boundary2} \\ 
 \left[ \left( P_0 + \varepsilon_0 \right) \xi 
 + \mu_{\rm n0} \mu_{i0} Y_{{\rm n}i} \, z_i \right]_{r=R_0}
&=& \left[ \left( P_0 + \varepsilon_0 \right) \xi_{\rm c} \right]_{r=R_0}. 
\label{boundary3}     
\end{eqnarray}
With Eq. (\ref{boundary1}), 
the equalities (\ref{boundary2}) and (\ref{boundary3}) 
can be written:
\begin{eqnarray}
\left[ \delta P  -\delta P_{\rm c} \right]_{r=R_0} &=& 0,
\label{boundary4} \\
\mu_{i0} Y_{{\rm n}i} \, z_i \mid_{r=R_0} &=& 0.
\label{boundary5}     
\end{eqnarray}
Eqs. (\ref{pc1}), (\ref{lim}), 
(\ref{boundary1}), (\ref{boundary4}), and (\ref{boundary5})
cover all boundary conditions that are to be imposed 
on Eqs. (\ref{sfleq5}) and (\ref{pulseq}) 
in order to find the frequency spectrum 
for the present neutron star model.

\section{Sound waves in superfluid mixtures}
Before discussing the numerical solutions 
to the pulsation equations (\ref{sfleq5}) and (\ref{pulseq}), 
let us analyze sound waves in superfluid neutron stars. 
One would expect the numerical solutions to resemble a
``plane'' sound wave when 
the number of nodes $N$ of the eigenfunctions $\xi$ and $z_i$ is large,
so that the wave number is large, $k \sim N/R \gg 1/R$.
Taking into account the estimate $\omega /k \sim u$, 
where $u$ is the sound velocity, 
we see that the eigenfrequencies of such ``sound-like"
modes must obey the inequality
\begin{equation}
\omega \gg u/R.
\label{inequality}     
\end{equation}
%
\begin{figure}
\setlength{\unitlength}{1mm}
\leavevmode
\hskip  0mm
\includegraphics[width=120mm,bb=18  145  562  690,clip]{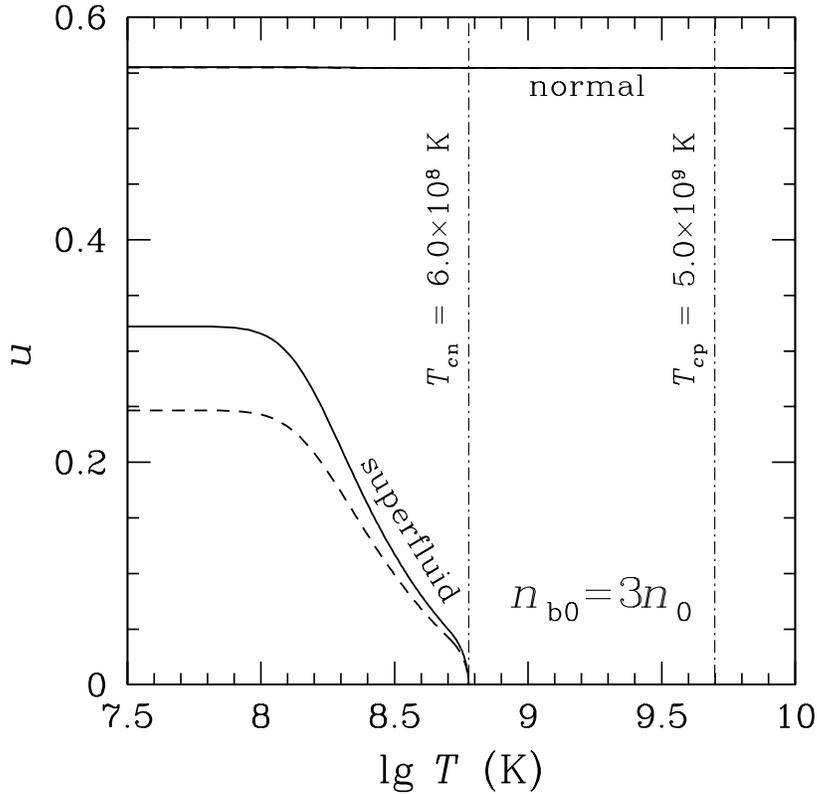}
\caption{Sound velocities $u_{1,2}$ (in units of $c$) 
as a function of temperature $T$ for two models of
the nucleon-nucleon potential: BJ v6 (solid lines) 
and Reid v6 (dashed lines). 
The $u_1(T)$ and $u_2(T)$ curves 
are marked as ``normal" and ``superfluid", respectively. 
The neutron and proton critical temperatures
are indicated by vertical dot-and-dashed lines. 
The baryon number density is $n_{\rm b0}=3 n_0$=0.48 fm$^{-3}$.}
\label{1}   
\end{figure}
We now simplify Eqs. (\ref{sfleq5}) and (\ref{pulseq}) 
to the case of short wavelength oscillations. 
Since the characteristic scale $R$ 
of variation of the equilibrium parameters 
(marked with the subscript ``0") 
is much larger than the characteristic scale $1/k$ 
of the variation of the eigenfunctions, 
we can neglect the spatial derivatives 
of the ``equilibrium" quantities 
and rewrite the pulsation equations as
\begin{eqnarray}
\mu_{\rm n0} \, {\rm e}^{\lambda_0-\nu_0} \, 
\omega^2 \, \left( z_{\rm n} +\xi \right) &=& 
{\partial \delta \mu_{\rm n} \over \partial r}, 
\label{soundeq1} \\
-{\rm e}^{\lambda_0-\nu_0} \, \omega^2 \, \widetilde{T}_0^{\, 1} &=& 
{\partial \delta P \over \partial r}.
\label{soundeq2}     
\end{eqnarray}
Under the assumption of a frozen nuclear composition, 
the functions $\delta P$ and $\delta \mu_{\rm n}$ 
are defined by Eqs. (\ref{dp2}) and (\ref{dmu2}), 
as before, but now we have
\begin{equation}
\Phi = {\partial \xi \over \partial r}, \quad \quad 
\Psi = {\mu_{k0} Y_{{\rm n}k} \over n_{\rm b0}} \, {\partial z_k \over \partial 
r}.
\label{phipsi}     
\end{equation}
The functions $\xi$ and $z_i$ can be presented in the form
\begin{equation}
\xi = \xi_0(r) \,{\rm e}^{i(kr-\omega t)}, 
\quad \quad z_i=z_{i0}(r) \, {\rm e}^{i(kr-\omega t)}.
\label{solution}     
\end{equation}
The derivatives of the slowly varying 
functions $\xi_0(r)$ and $z_{i0}(r)$ can be ignored. 
By substituting the expressions (\ref{solution})
into Eqs. ({\ref{soundeq1}) and (\ref{soundeq2}), 
one can find from the compatibility condition of 
the resulting set of equations,
a biquadratic equation for the local sound velocity 
$u={\rm e}^{(\lambda_0-\nu_0)/2} \, \omega/k$ :
\begin{equation}
y \, u^4 + \left[ {P_0 \over \mu_{\rm n0} n_{\rm b0}} \,\left(
\beta_1 - \gamma_1 - \gamma_1 y \right) + \gamma_2-\beta_2   \right] u^2
+ {P_0 \over \mu_{\rm n0} n_{\rm b0}} \, 
\left( \beta_2 \gamma_1 - \beta_1 \gamma_2 \right) =0.
\label{bi2}     
\end{equation}
This equation has two nontrivial solutions 
for two possible sound velocities (see Andersson and Comer 2001a for a similar discussion). 
The dimensionless parameter $y$ is defined as
\begin{equation}
y = {Y_{\rm pp} \, n_{\rm b0} \over \mu_{\rm n0} \, 
\left(Y_{\rm nn} Y_{\rm pp} - Y_{\rm np} Y_{\rm pn} \right)} -1.
\label{y}     
\end{equation}
At $T \rightarrow T_{c {\rm n}}$, we have 
$Y_{\rm nn}$, $Y_{\rm np}$, $Y_{\rm pn}$ $\rightarrow 0$, 
hence, 
$y \approx n_{\rm b0}/(\mu_{\rm n0} \, Y_{\rm nn}) \rightarrow \infty$.
In this case the roots of Eq. (\ref{bi2}) 
are approximately given by
\begin{equation}
u_1 \approx \sqrt{ {P_0 \gamma_1 \over \mu_{\rm n0} n_{\rm b0} } }, \quad \quad
u_2 \approx \sqrt{ { \mu_{\rm n0} \, Y_{\rm nn} \over n_{\rm b0} \, \gamma_1} \, 
\left( \beta_2 \gamma_1 - \beta_1 \gamma_2 \right)}.
\label{bi3}     
\end{equation}
The first root describes the velocity of perturbations 
similar to the familiar sound propagating 
through a medium with non-superfluid neutrons. 
The second root indicates the existence of 
an additional pulsation mode specific to superfluid matter. 
For the second mode to be stable 
the condition 
$\beta_2 \gamma_1 \geq \beta_1 \gamma_2$ 
must be fulfilled. 
The second pulsation mode vanishes at 
$T>T_{c{\rm n}}$ ($Y_{\rm nn}=Y_{\rm np}=Y_{\rm pn}=0$), 
while the velocity of the first mode 
is still defined by Eq. (\ref{bi3}). 
In that case, the first mode is just the usual sound.

The results of a numerical solution of Eq. (\ref{bi2}) 
for matter with baryon number density $n_{\rm b0}=3 n_{0}$ 
are presented in Fig. 1
($n_0=0.16$ fm$^{-3}$ is the baryon 
number density in atomic nuclei). 
In determining these data
we used critical temperatures for neutrons and protons equal to 
$T_{c{\rm n}}=6 \times 10^8$ K and $T_{c {\rm p}}= 5 \times 10^9$ K,
respectively, and employed the equation of state 
of Heiselberg and Hjorth-Jensen (1999) 
to calculate the thermodynamic parameters and their derivatives. 
The velocities $u_{1,2}$ (in units of $c$) 
are plotted as a function of temperature $T$ 
for two models of nucleon-nucleon potential: BJ v6 (solid lines) 
and Reid v6 (dashed lines). 
Note that, the choice of the model potential
determines the entrainment matrix $\rho_{ik}$ 
and, hence, the matrix  $Y_{ik}$
(see the paper by Gusakov and Haensel 2005; the 
microphysics is described 
by Jackson et al. 1982). 
The $u_2(T)$ curves are marked  
``superfluid" and the $u_1(T)$ curves are marked ``normal". 
One can see that the sound velocity
$u_1(T)$ is practically insensitive to the model potential chosen: 
the solid and dashed lines in the figure coincide.

The analysis of Fig. 1 shows that the results 
of a numerical solution of Eq. (\ref{bi2}) 
are generally consistent with the above conclusions. 
We would like to stress that the sound velocity 
$u_1$ does not significantly differ from that calculated 
from Eq. (\ref{bi3}) even at $T \ll T_{c {\rm n}}$. 
It is also important that 
the velocity of the second mode $u_2$ becomes 
comparable 
to the velocity $u_1$ at low temperatures, 
in contrast to the case of pure helium II. 
At $T \la 0.5 T_{c{\rm n}}$ the velocity $u_2$
rapidly approaches its asymptotic value $u_{2}(T=0)$.

Let us discuss briefly sound in beta-equilibrated matter. 
It is easy to verify that all the formulas derived in this section 
remain valid, provided that the thermodynamic parameters 
and their derivatives are considered as functions of {\it only}
the baryon number density $n_{\rm b0}$ and entropy $S_0$. 
(We recall that the electron number density $n_{\rm e0}$ 
is not an independent variable in this case; 
it is  determined by the beta-equilibrium condition.) 
In particular, the adiabatic index is now: 
$\gamma_1=(n_{\rm b0}/P_0) \,\, \partial P(n_{\rm b0}, 
x_{\rm s0})/\partial n_{\rm b0}$.

Using $n_{\rm b0}$ and $S_0$ as independent variables 
instead of $n_{\rm b0}$ and $x_{\rm s0}$ in the functions
$\gamma_1$ and $\gamma_2$ and expressing the derivatives 
$\partial \mu_{\rm n}(n_{\rm b0}, S_0) /\partial n_{\rm b0}$ and
$\partial \mu_{\rm n}(n_{\rm b0}, S_0) /\partial S_0$ 
in Eq. (\ref{bi2}) with the help of 
the Gibbs-Duhem relation, 
$\dd P=S_0 \, \dd T + n_{\rm b0} \, \dd \mu_{\rm n}$,
we get
\begin{equation}
y \, u^4 - {1 \over \mu_{\rm n0} n_{\rm b0}} \, 
\left( S_0^2 \, {\partial T \over \partial S_0} + y n_{\rm b0} \, 
{\partial P \over \partial n_{\rm b0}} + S_0 y \, 
{\partial P \over \partial S_0} \right) u^2 
+ {S_0^2 \over \mu_{\rm n0}^2 n_{\rm b0}} \, 
\left( {\partial P \over \partial n_{\rm b0}} \, {\partial T \over \partial S_0}
-{\partial P \over \partial S_0} \, 
{\partial T \over \partial n_{\rm b0}} \right) =0.
\label{bi4}     
\end{equation}
An approximate solution to this equation 
can be easily found if we keep in mind 
that we always have $u_1 \gg u_2$:
%
\begin{equation}
u_1 \approx \sqrt{{1 \over \mu_{\rm n0}} 
\, {\partial P \over \partial n_{\rm b0}}},
\quad \quad
u_2 \approx \sqrt{{S_0^2 \over \mu_{\rm n0} n_{\rm b0} \, y} \, 
{\partial T \over \partial S_0}}.
\label{bi5}     
\end{equation}
%
\begin{figure}
\setlength{\unitlength}{1mm}
\leavevmode
\hskip  0mm
\includegraphics[width=120mm,bb=18  145  562  690,clip]{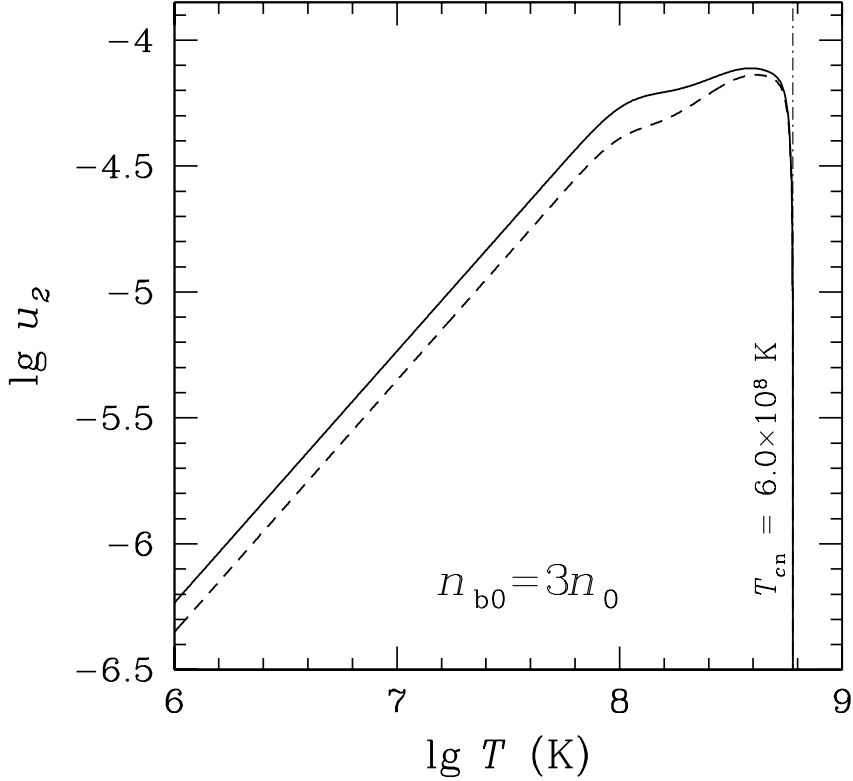}
\caption{The velocity of the second sound $u_2$ (in units of $c$) 
in beta-equilibrated matter as a function 
of temperature $T$ for the same models of the nucleon-nucleon potential, 
neutron and proton superfluidity, 
and baryon number density $n_{\rm b0}$ as in Fig.~1.}
\label{2}   
\end{figure}
Again, the first root describes the velocity of sound
in non-superfluid matter (the first sound) 
and the second root describes the velocity of the so-called second sound. 
It should be noted that the first sound cannot, 
in fact, propagate with the velocity $u_1$ 
defined by Eq. (\ref{bi5}), 
because this velocity is so high that 
no beta-equilibrium can exist in such a wave. 
If we use Eq. (\ref{bi5}) to describe the sound 
in a one-component liquid, the expression for $u_2$ 
in the non-relativistic limit will coincide with that 
for the second sound in liquid helium II 
(see, e.g., Khalatnikov 1989).

The function $u_2(T)$ 
for matter with  baryon number density 
$n_{\rm b0}=3 n_0$ is shown in Fig. 2. 
We used the same models of superfluidity and the nucleon-nucleon 
potentials and the same equation of state 
as in the discussion of sound 
in matter with frozen nuclear composition. 
The speed of sound was calculated numerically.
While doing the calculations we used the formula 
$C = T \, \partial S/\partial T$,  
where $C$ is the heat capacity of superfluid matter 
(an expression for $C$ can be found, 
e.g., in Yakovlev et al. 1999).

It follows from Eq. (\ref{bi5}) and Fig. 2 
that the velocity $u_2$ goes to zero 
at both $T=T_{c{\rm n}}$ and $T = 0$. 
However, beta-processes are so suppressed 
at low temperatures that the second 
sound will not be able to propagate, 
because matter cannot approach 
beta-equilibrium  on a timescale
comparable with the pulsation period. 
Therefore, the  second 
sound can only exist in a range of temperatures near $T \la T_{c{\rm n}}$.

To conclude, three types of sound waves can exist in superfluid npe-matter.
The speed of two of them is so high
that they 
propagate in matter with a frozen nuclear composition, 
while the waves of the third type 
can 
exist
only in beta-equilibrated matter 
at temperatures in the vicinity of 
the neutron critical temperature $T_{c {\rm n}}$.

\section{Results for radial pulsations}
Let us now discuss the solutions 
to the pulsation equations (\ref{sfleq5}) and (\ref{pulseq}). 
We have integrated the equations in a standard way, 
using the Runge-Kutta method. 
We employed the 
equation of state  
of Negele and Vautherin (1973) in the stellar crust 
and that of Heiselberg and Hjorth-Jensen (1999) in the core. 
The latter is a convenient analytical approximation to the 
equation of state proposed by Akmal and Pandharipande (1997). 
For this equation of state,
the most massive stable neutron 
star has central density
$\rho_{\rm c}=2.76 \times 10^{15}$ g~cm$^{-3}$,
circumferential radius $R=10.3$ km, and 
 gravitational mass $M = M_{\rm max} = 1.92 \, M_\odot$.
The powerful direct Urca process 
of neutrino emission is open in the 
core of a star of mass $M > 1.83\, M_\odot$.
When calculating the matrix $Y_{ik}$ we have used 
the model BJ v6 of nucleon-nucleon potential 
(see Section 5).

For illustration, we consider a neutron star model  
with mass $M =1.4 M_{\odot}$ 
($R=12.17$ km, $\rho_{\rm c}=9.26 \times 10^{14}$ g cm$^{-3}$). 
For such a star the crust-core interface 
is at $R_{\rm cc}=10.88$ km. 
%
%
The frequencies of the first three modes of radial pulsations 
of a non-superfluid star with this mass are 
$\omega_1 = 1.703 \times 10^4$~s$^{-1}$, 
$\omega_2 = 4.081 \times 10^4$~s$^{-1}$, and 
$\omega_3 = 5.732 \times 10^4$~s$^{-1}$.

To reduce the number of factors affecting the pulsation spectrum, we 
consider a simplified superfluidity model 
in which the critical red-shifted temperatures of nucleons 
do not vary with the density and are equal to 
$T_{c {\rm n}}^{\infty} \equiv T_{c{\rm n}} 
{\rm e}^{\nu_0/2} = 6 \times 10^8$~K and
$T_{c {\rm p}}^{\infty} \equiv T_{c{\rm p}} 
{\rm e}^{\nu_0/2} = 5 \times 10^9$~K. 
Consequently, superfluid matter
is contained in the stellar core: $R_0=R_{\rm cc}$. 
This means that the boundary at $R_0$ is
``attached'' to matter and, for example,
is independent of temperature variations.
(Note that, in the more general case of density dependent profiles 
of critical temperatures, 
the superfluid-normal boundary {\it can} 
depend on $T$ and temperature perturbations.)
Numerical tests have shown that the approximation 
of critical temperatures $T_{c {\rm n,p}}^{\infty}$ as   
constant throughout the core 
describes  reality well 
if these temperatures smoothly depend on the density. 
This is consistent with the predictions of some 
microscopic models of nucleon pairing known in literature
(see, e.g., Yakovlev et al. 1999).

\begin{figure}
\setlength{\unitlength}{1mm}
\leavevmode
\hskip  0mm
\includegraphics[width=120mm,bb=18  145  562  690,clip]{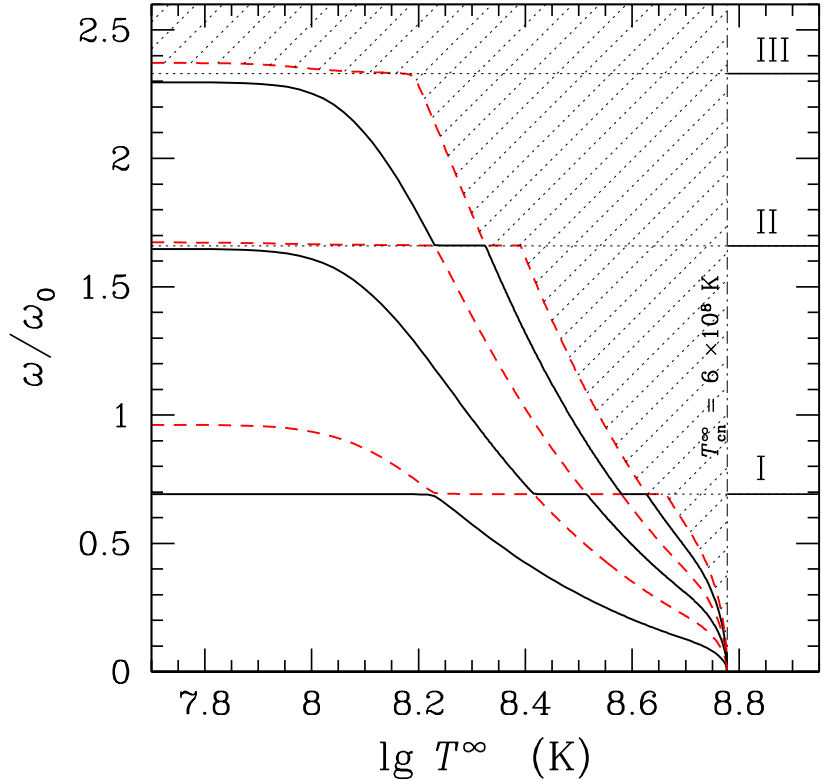}
\caption
{The pulsation eigenfrequencies $\omega$  
(in units of $\omega_0=c/R$) 
of a neutron star as a function of 
the red-shifted core temperature $T^{\infty}$. 
The neutron critical 
temperature $T_{c{\rm n}}^{\infty}$ 
is indicated by the vertical dot-and-dashed line; 
the horizontal dotted lines 
changing into solid lines at 
$T^{\infty}> T^{\infty}_{c{\rm n}}$ 
indicate the first three eigenfrequencies (I, II, III) 
of a non-superfluid star. 
No spectrum was plotted in the shaded region. 
The dashed curves correspond to ``superfluid" modes 
at $T^{\infty} \ll T^{\infty}_{c{\rm n}}$ and 
the solid curves correspond to ``normal" modes 
at $T^{\infty} \ll T^{\infty}_{c{\rm n}}$ (see text).}
\label{3}   
\end{figure}
Fig. 3 shows the dependence of the pulsation eigenfrequencies 
$\omega$  on the 
red-shifted temperature $T^{\infty} \equiv T_0 {\rm e}^{\nu_0/2}$  
(we recall that the superfluid core 
is isothermal, in accordance with Eq. \ref{gradT}). 
The vertical dot-and-dashed line indicates 
the neutron critical temperature $T^{\infty}_{c {\rm n}}$. 
The horizontal dotted lines show the first three eigenfrequencies 
$\omega_1$, $\omega_2$, and $\omega_3$  
for a non-superfluid star. 
No attempt to determine the spectrum in the shaded region was made. 
At $T^{\infty} > T^{\infty}_{c {\rm n}}$, 
the star pulsates 
as a normal fluid
(no matter whether the protons are paired or not).
Hence, the spectrum contains only normal, 
temperature-independent pulsation modes 
(the first three modes I, II, and III 
are shown as solid lines). 
At $T^{\infty} \la 0.1 T^{\infty}_{c {\rm n}}$, 
a pulsating star can be described in 
the zero-temperature approximation. 
The spectrum of a cold superfluid star is doubled, 
as compared with that of a normal star 
(see Comer et al. 1999). 
In addition to ``normal" pulsation modes, 
whose eigenfrequencies are close to those for
a non-superfluid star (solid lines), 
the spectrum contains specific ``superfluid" modes (dashed lines). 
Note that, the first ``superfluid'' mode is quite 
different from the ``normal'' one but the second and 
third ``superfluid'' modes are already sufficiently 
close to their ``normal'' counterparts (see Fig. 3).

As the temperature increases, 
starting from approximately $T^{\infty} \sim 10^8$ K, 
the frequency of each mode begins to decrease. 
When a mode reaches one of the horizontal dotted lines, 
it changes behavior and becomes temperature-independent, 
{\it imitating} the behavior of one of the non-superfluid modes. 
As the temperature rises further, 
the frequency of the higher mode approaches that of the mode in question, 
which in turn begins to decrease again 
(see avoiding crossings in Fig. 3). 
As a result, the two different 
modes of the spectrum will never intersect.
One can conclude that a given mode may behave either as ``superfluid'' 
or ``normal'' with increasing temperature.

The behavior of the frequency spectrum 
at temperatures close to $T^{\infty}_{c {\rm n}}$ is of particular interest. 
It is clear from Fig. 3 that the frequency of 
any mode goes to zero at $T^{\infty} = T^{\infty}_{c{\rm n}}$. 
This is not surprising if we keep in mind that 
high order pulsation modes represent sound-like waves (see Section 5), 
and that the frequency of the ``superfluid" sound 
also goes to zero at the transition point 
into the superfluid state (Fig. 1). 
It might seem that the spectrum does not contain 
eigenfrequencies of non-superfluid stars 
at the transition point when all neutrons in the star are normal. 
However, this is not the case. 
The point is that at $T^{\infty} \rightarrow T^{\infty}_{c{\rm n}}$, 
the number of modes with frequencies in any given interval, 
say, $[0, \omega_1]$, becomes infinitely large. 
As a result, at any temperature $T^{\infty}$ 
and any eigenfrequency of a normal star, 
there is a mode which is temporarily ``normal-like", 
i.e., it has the same frequency as in the normal fluid.

Since the temperature of neutron stars changes with time, 
it would be interesting to discuss how the pulsation frequencies 
vary with time. 
Suppose that the pulsation energy is much 
lower than the thermal energy. 
We can then neglect the star heating due to the conversion of 
pulsation energy into heat 
(see Gusakov et al. 2005 for details). 
The star will cool down, 
and to determine the dependence of the
internal temperature 
$T^{\infty}$ on time $t$ one should use 
the cooling theory of superfluid neutron stars 
(see, e.g., Yakovlev et al. 1999, 
Yakovlev and Pethick 2004).

Since the direct Urca process is forbidden for 
the chosen neutron star model, 
the main cooling mechanisms 
(at $T^{\infty} < T^{\infty}_{c{\rm n}}$) 
will be neutrino emission due to Cooper pairing of neutrons, 
neutron-neutron bremsstrahlung, 
and the photon emission from the stellar surface. 
One can easily find the function $T^{\infty}(t)$ 
by solving the thermal balance equation (see, e.g., Yakovlev et al. 1999) 
under the assumption that 
the stellar core is isothermal.
If the dependencies $\omega(T^{\infty})$ and $T^{\infty}(t)$ are known,
it is possible to plot the frequency spectrum
$\omega$ as a function of time $t$ (Fig. 4). 
Here the time (in units of $10^3$ years) 
is counted from the moment of neutron superfluidity onset 
(at $T^{\infty} =T^{\infty}_{c {\rm n}}$). 

\begin{figure}
\setlength{\unitlength}{1mm}
\leavevmode
\hskip  0mm
\includegraphics[width=120mm,bb=18  145  562  690,clip]{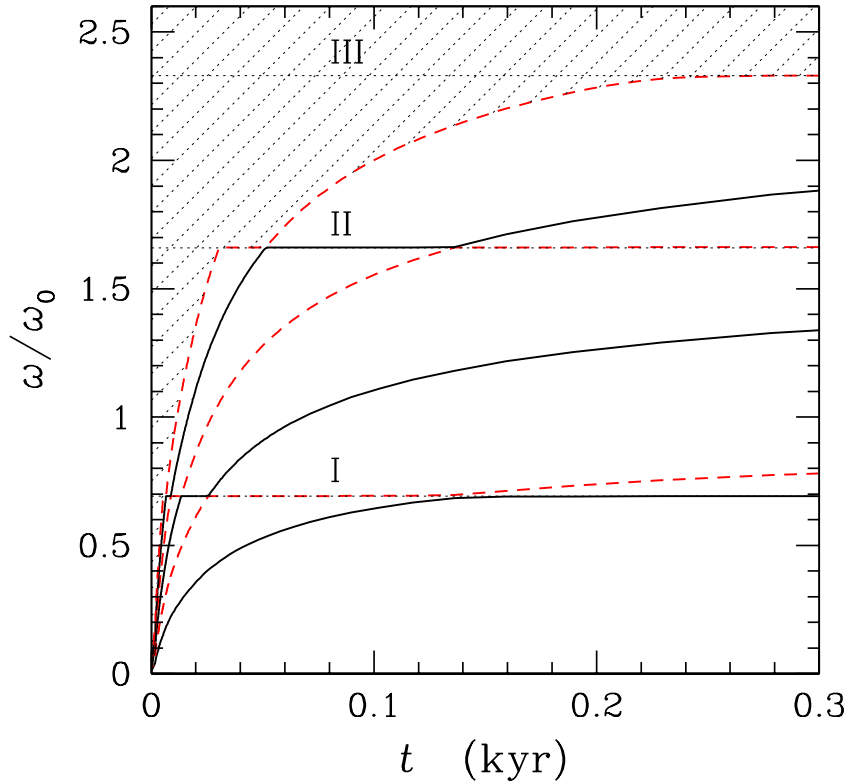}
\caption{The dependence of the pulsation spectrum of 
a superfluid neutron star on time $t$, 
counted from the moment of neutron superfluidity onset 
(at $T^{\infty}=T^{\infty}_{c{\rm n}}$). 
The time is in units of $10^3$ years. 
Notations are the same as in Fig. 3.}
\label{4}   
\end{figure}
The analysis of Fig. 4 shows 
a significant change in the pulsation spectrum
for 20 years after superfluidity turns on. This is 
 associated with the highly efficient
Cooper pairing neutrino emission 
(the detailed discussion of this process and its influence 
on the neutron star cooling is given by Gusakov et al. 2004).
For example, the frequency of the third ``superfluid"
mode changes during this period of time from $0$ to the eigenfrequency 
$\omega_2$ of a non-superfluid star. 
The Cooper pairing neutrino emission process 
quickly becomes weaker with time, 
the cooling slows down and 
the variation in $\omega(t)$ becomes smoother. 
We would like to emphasize that 
the fast change of the pulsation frequencies 
for the first few dozens of years 
is due to the  high critical temperature of neutrons, 
$T^{\infty}_{c{\rm n}}=6 \times 10^8$ K. 
We could make the $\omega(t)$ dependence less dramatic 
by choosing lower critical temperatures.

\section{Summary}

The aim of the present study was to analyze radial pulsations 
of superfluid neutron stars at finite core temperatures.
We used the equations for one-component superfluid 
hydrodynamics suggested by Son (2001), rewritten in terms of 
more convenient variables, 
and extended to the case of superfluid mixtures in General Relativity. 
A simple model of npe-matter was employed 
to show that a necessary condition 
for a star to be at hydrostatic and beta-equilibrium
is constancy of the red-shifted temperature 
in the region of the star where the neutrons are superfluid: 
$T {\rm e}^{\nu_0/2}={\rm constant}$. 
Proton superfluidity does {\it not} 
impose any restrictions on the temperature, 
because protons are ``coupled" with
normal electrons by electromagnetic forces 
and behave as a normal fluid, no matter whether they 
are superfluid or not.

The hydrodynamics of superfluid mixtures 
was applied to investigate radial pulsations of neutron stars. 
It was assumed that the crust is non-superfluid, 
and neutrons and protons have red-shifted critical temperatures,
which are constant throughout the core. 
The set of equations we have derived 
describes radial pulsations of superfluid stellar matter.

We have found the short wavelength solutions to this 
set of equations, representing sound waves in
superfluid neutron star matter. 
The dependence of the speed of sound on the stellar temperature
was examined in two limiting pulsation regimes: 
(1) in beta-equilibrated pulsating matter
and (2) in pulsating matter with frozen nuclear composition.
It was shown that three different kinds of sound waves 
may  in principle exist, 
two of them propagate in the matter with 
frozen nuclear composition and one can exist only 
in beta-equilibrium. 
While the speeds of the former sound waves are comparable 
to each other (see Fig. 1) 
and to the speed of sound in the usual non-superfluid matter, 
the speed of the latter is 4--5 orders of magnitude lower (see Fig. 2); 
it can be excited only at temperatures $T$ close to $T_{c{\rm n}}$.

Generally, the pulsation equations were solved numerically, 
and the results show that the finite internal temperatures 
strongly affect the pulsation spectrum in the range of
$T \sim (0.1-1) T_{c{\rm n}}$ (see Fig. 3). 
The frequency of any pulsation mode in this range 
decreases with increasing temperature. 
However, when the mode reaches one of the eigenfrequencies 
of a non-pulsating star, 
it becomes temperature independent for a while.
One may say that it begins to {\it mimic} the behavior of 
a non-superfluid mode. 
At $T \rightarrow T_{c {\rm n}}$, 
all superfluid eigenfrequencies tend to zero. 
At $T \la 0.1 T_{c{\rm n}}$, 
the pulsation spectrum is similar to that calculated 
in the zero temperature approximation.

In addition to the analysis of 
the temperature dependence of the pulsation spectrum, 
we discuss the temporal evolution of the eigenfrequencies 
during the star cooling (Fig. 4). 
In our analysis, we use the standard cooling theory 
of superfluid neutron stars (see, e.g., Yakovlev et al. 1999). 
The calculation shows that essential 
changes (within the present model) in 
the pulsation eigenfrequencies occur 
for the first 20 years 
following the moment of neutron superfluidity onset. 
This rather short (for the cooling theory) 
period of time is associated with the fast cooling 
due to the effective Cooper pairing neutrino emission process.
It will be even shorter 
if the powerful direct Urca process operates in the stellar core.

The consideration of the problem 
presented here is based on a simplified model.
In particular, we discuss only the simplest case of radial pulsations 
and assume critical temperatures of nucleons that are constant throughout the core.
However, it would be important (and interesting) 
to understand how finite internal temperatures affect 
the frequency spectrum of non-radial pulsations 
and how the results would change 
if we analyzed more realistic density profiles 
for the critical temperatures. 
Finally, in a more realistic approach  
one should take into account 
$^1S_0$ neutron pairing in the stellar crust 
and more accurately treat the physics of the crust, 
especially if one deals with pulsation modes localized
in the outer layers of the star. 
In spite of the considerable simplification of 
the problem discussed in this paper, 
we conclude that finite internal temperatures 
{\it significantly affect} 
the pulsation spectrum of 
not too cold superfluid neutron stars. 
Moreover, the pulsation frequencies 
can change dramatically 
for a period of several dozens of years, an effect that may potentially be 
observable.

\section*{Acknowledgments}
The authors are grateful to D.G. Yakovlev 
for discussion. 
One of the authors (MEG) also acknowledges
excellent working conditions 
at the School of Mathematics (University of Southampton) 
in Southampton, where part of this study was done.

\noindent This research was supported
by RFBR (grants 05-02-16245 and 05-02-22003),
the Russian Leading Science School (grant 9879.2006.2),
INTAS YSF (grant 03-55-2397), 
and by the Russian Science Support Foundation.

\bsp 

\end{document}